\documentclass[aps,pra,twocolumn,amsmath,amssymb,showpacs]{revtex4}
\usepackage{amsmath}
\usepackage{graphicx}
\usepackage{epsfig}
\usepackage{amsfonts}

\begin{document}
\title{Spectrum of single-photon emission and scattering in cavity optomechanics}
\author{Jie-Qiao Liao}
\affiliation{Department of Physics and Institute of Theoretical
Physics, The Chinese University of Hong Kong, Shatin, Hong Kong
Special Administrative Region, People's Republic of China}
\author{H. K. Cheung}
\affiliation{Department of Physics and Institute of Theoretical
Physics, The Chinese University of Hong Kong, Shatin, Hong Kong
Special Administrative Region, People's Republic of China}
\author{C. K. Law}
\affiliation{Department of Physics and Institute of Theoretical
Physics, The Chinese University of Hong Kong, Shatin, Hong Kong
Special Administrative Region, People's Republic of China}
\date{\today}

\begin{abstract}
We present an analytic solution describing the quantum state of a
single photon after interacting with a moving mirror in a cavity.
This includes situations when the photon is initially stored in a
cavity mode as well as when the photon is injected into the cavity.
In addition, we obtain the spectrum of the output photon in the
resolved-sideband limit, which reveals spectral features of the
single-photon strong-coupling regime in this system. We also clarify
the conditions under which the phonon sidebands are visible and the
photon-state frequency shift can be resolved.
\end{abstract}
\maketitle

Cavity optomechanics~\cite{Kippenberg2008,Marquardt2009} is a new
frontier for exploring the coherent coupling via radiation pressure
between electromagnetic and mechanical degrees of freedom. Several
recent experimental systems in cavity
optomechanics~\cite{Gupta2007,Brennecke2008,Eichenfield2009}
approached the single-photon strong-coupling regime, in which the
radiation pressure from a single photon can produce observable
effects. Such a strong coupling is important to realize various
proposals of generating macroscopic superposition states for testing
quantum theory~\cite{Mancini1997,Bose1997,Bouwmeester2003}. In
parallel with the progress in experiments, some theoretical
investigations have also
begun~\cite{Rabl2011,Nunnenkamp2011,Chenp2011} to explore the
effects of radiation-pressure coupling at few- and even
single-photon levels. For example, Rabl~\cite{Rabl2011}, and
Nunnenkamp and coworkers~\cite{Nunnenkamp2011} have independently
studied the statistics of cavity photons in the single-photon
strong-coupling regime when the cavity is weakly driven by a
continuous-wave laser.

Physically, in the single-photon strong-coupling regime, the
mirror's displacement induced by a single photon can significantly
affect the cavity field, leading to some observable features in the
spectrum of single-photon emission and scattering. Thus, a natural
question is how these spectra may characterize the single-photon
strong-coupling regime. In this Brief Report we answer the question
by calculating analytically the spectrum of single-photon emission
and scattering in a cavity optomechanical system. In particular, we
indicate the relation between the photon spectral characteristic and
the interaction strength $g$ [cf. Eq.~(\ref{optocavityH})] of
radiation pressure per photon.

\begin{figure}[tbp]
\center
\includegraphics[bb=80 467 328 758, width=3.3 in]{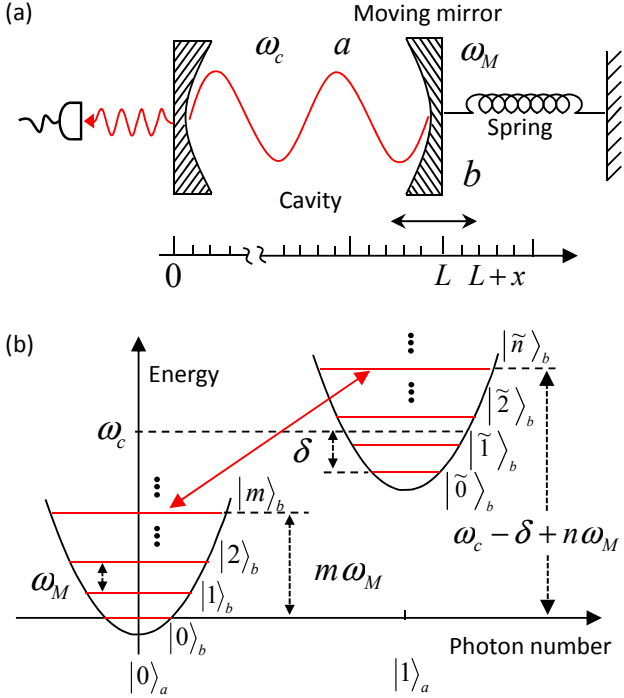}
\caption{(Color online) (a) Schematic diagram of a Fabry-Perot-type
optomechanical cavity formed by a fixed end mirror and a moving end
mirror. (b) The energy-level structure of the optomechanical cavity
(limited in the zero- and one-photon subspaces). Here we denote
$\vert\tilde{n}(0)\rangle_{b}=\vert n\rangle_{b}$ and
$\vert\tilde{n}(1)\rangle_{b}=\vert\tilde{n}\rangle_{b}$.}\label{setup}
\end{figure}

To begin with, we specify the system of a Fabry-P\'{e}rot-type
optomechanical cavity formed by a fixed end mirror and a moving end
mirror [Fig.~\ref{setup}(a)]. The cavity field and the mirror motion
are coupled to each other via radiation pressure. The Hamiltonian of
the optomechanical cavity is
\begin{equation}
H_{\textrm{opc}}=\hbar\omega_{c}a^{\dagger}a+\hbar\omega_{M}
b^{\dagger}b-\hbar ga^{\dagger}a(b^{\dagger}+b),\label{optocavityH}
\end{equation}
where $a$ $(a^{\dagger})$ and $b$ $(b^{\dagger})$ are, respectively,
the annihilation (creation) operators of the electromagnetical and
mechanical modes, with the respective resonant frequencies
$\omega_{c}$ and $\omega_{M}$. The parameter
$g=\omega_{c}x_{\textrm{zpf}}/L$ is the single-photon coupling
strength of the radiation pressure between the cavity and the
mirror, where $x_{\textrm{zpf}}=\sqrt{\hbar/(2M\omega_{M})}$ is the
zero-point fluctuation of the mirror with mass $M$, and $L$ is the
rest length of the cavity.

Let us denote the harmonic-oscillator number states for the cavity
and the mirror as $|m\rangle_{a}$ and $|n\rangle_{b}$ ($m,n=0,1,2,
\cdots$), respectively, then the eigensystem of
Hamiltonian~(\ref{optocavityH}) can be written as
\begin{equation}
H_{\textrm{opc}}|m\rangle_{a}|\tilde{n}(m)\rangle_{b}
=\hbar(m\omega_{c}+n\omega_{M}-m^{2}\delta
)|m\rangle_{a}|\tilde{n}(m)\rangle_{b}
\end{equation}
for $n,m=0,1,2,\cdots$, where $\delta=g^{2}/\omega_{M}$ is the
photon-state frequency shift caused by a single-photon radiation
pressure. The state
\begin{equation}
|\tilde{n}(m)\rangle_{b}\equiv\exp[m\beta_{0}(b^{\dagger}-b)]|n\rangle_{b}\label{eq3}
\end{equation}
is an $m$-photon displaced number state~\cite{Oliveira1990} with
$\beta_{0}=g/\omega_{M}$. Particularly, $|\tilde{n}(1)\rangle_{b}$
is the single-photon displaced number state; and
$|\tilde{n}(0)\rangle_{b}$ is the zero-photon displaced number
state, which is the same as the harmonic-oscillator number state
$|n\rangle_b$ by Eq.~(\ref{eq3}). For convenience, the energy-level
structure of the optomechanical cavity in the zero- and one-photon
cases is shown in Fig.~\ref{setup}(b).

In optomechanical systems, the decay rate $\gamma_m$ of the mirror
motion can be much smaller than the cavity-field's decay rate
$\gamma_c$, then during the time interval $1/\gamma_c\ll t\ll
1/\gamma_m$, the damping of the mirror motion is negligible. In the
following we merely consider the dissipation of the cavity field by
coupling it to continuous outside fields so that the full
Hamiltonian is
\begin{equation}
H=H_{\textrm{opc}}+\int_{0}^{\infty}\hbar\omega_{k}c_{k}^{\dagger}c_{k}dk
+\hbar\xi\int_{0}^{\infty}(c_{k}^{\dagger}a+a^{\dagger}c_{k})dk,\label{hamiltonian-rot}
\end{equation}
where $c_{k}$ ($c_{k}^{\dagger }$) is the annihilation (creation)
operator of the $k$th outside-field mode with frequency
$\omega_{k}$, and $\xi$ is the photon-hopping interaction strength.

Denoting $\vert\tilde{n}(1)\rangle_{b}=\vert \tilde{n}\rangle_{b}$
for concision, a general state in the single-photon subspace of the
total system can be expressed as
\begin{eqnarray}
|\varphi(t)\rangle&=&\sum_{n=0}^{\infty}A_{n}(t)\vert
1\rangle_{a}\vert\tilde{n}\rangle_{b}\vert
\emptyset\rangle\nonumber\\
&&+\sum_{n=0}^{\infty}\int_{0}^{\infty
}B_{n,k}(t)\vert0\rangle_{a}\vert n\rangle_{b}\vert 1_{k}\rangle
dk,\label{generalstate}
\end{eqnarray}
where $\vert
1\rangle_{a}\vert\tilde{n}\rangle_{b}\vert\emptyset\rangle$ stands
for the state with one photon in the cavity, no photon in these
outside fields, and the mirror in displaced number state
$|\tilde{n}\rangle_{b}$ (hereafter we call the single-photon
displaced number state as displaced number state when there is no
confusion), while $\vert 0\rangle_{a}\vert n\rangle_{b}\vert
1_{k}\rangle$ denotes the state with a vacuum cavity field, one
photon in the $k$th mode of these continuous fields, and the mirror
in number state $|n\rangle_{b}$. In Eq.~(\ref{generalstate}), we
have used the completeness relation of these displaced number states
$|\tilde{n}\rangle_{b}$ ($n=0,1,2,\cdots$). According to the
Schr\"{o}dinger equation, we can obtain equations of motion [in the
rotating picture with respect to
$\hbar\omega_{c}(a^{\dagger}a+\int_{0}^{\infty}
c_{k}^{\dagger}c_{k}dk)$] for probability amplitudes as
\begin{subequations}
\label{eqofmotion}
\begin{align}
\dot{A}_{m}(t)=&-i(m\omega_{M}-\delta)A_{m}(t)\nonumber\\
&-i\xi\sum_{n=0}^{\infty}\int_{0}^{\infty}\langle\tilde{m}\vert_{b}\vert
n\rangle_{b}B_{n,k}(t)dk,\\
\dot{B}_{m,k}(t)=&-i(m\omega_{M}+\Delta_{k})B_{m,k}(t)\nonumber\\&-i\xi
\sum_{n=0}^{\infty}\langle m\vert_{b}\vert
\tilde{n}\rangle_{b}A_{n}(t),
\end{align}
\end{subequations}
where $\Delta_{k}=\omega_{k}-\omega_{c}$ is the detuning of the
$k$th mode photon from the cavity frequency. In the following we
discuss the long-time solution and the spectrum of two cases.

\emph{Single-photon emission.} In this case, we assume that a single
photon is initially stored in the cavity mode, and such a photon
will be transmitted (or emitted) out of the cavity. We first
calculate the case where the initial state of the mirror is
harmonic-oscillator number state $|n_{0}\rangle_{b}$. Once the
solution for this case is found, the solution for general initial
states of the mirror can be obtained accordingly by superposition.
For the initial state
$|\varphi(0)\rangle=|1\rangle_{a}|n_{0}\rangle_{b}|\emptyset\rangle$,
we have
\begin{equation}
A_{m}(0)=\langle \tilde{m}\vert_{b}\vert n_{0}\rangle
_{b},\hspace{0.5 cm}B_{m,k}(0)=0.
\end{equation}
The transient
solution of Eq.~(\ref{eqofmotion}) can be obtained by the Laplace
transform method. The results are
\begin{subequations}
\begin{align}
A_{n_{0},m}(t)=&\langle \tilde{m}\vert _{b}\vert n_{0}\rangle
_{b}e^{-\left[\frac{\gamma_{c}}{2}
+i(m\omega_{M}-\delta)\right]t},\\
B_{n_{0},m,k}(t)=&\sqrt{\frac{\gamma_{c}}{2\pi}}\sum_{n=0}^{\infty}\frac{\langle
m\vert_{b}|\tilde{n}\rangle_{b}\langle \tilde{n}|_{b}\vert
n_{0}\rangle_{b}}
{\Delta_{k}+\delta-(n-m)\omega_{M}+i\frac{\gamma_{c}}{2}}\nonumber\\
&\times\left(e^{-i(m\omega_{M}+\Delta_{k})t}
-e^{-\left[\frac{\gamma_{c}}{2}+i(n\omega_{M}-\delta)\right]t}\right),
\end{align}
\end{subequations}
where $\gamma_{c}=2\pi\xi^{2}$ is the cavity-field decay rate.
Notice that the subscript $n_{0}$ in $A_{n_{0},m}(t)$ and
$B_{n_{0},m,k}(t)$ is added to mark the mirror's initial state
$|n_{0}\rangle_{b}$. In the long-time limit, we have
$A_{n_{0},m}(\infty)=0$ and
\begin{equation}
B_{n_{0},m,k}(\infty)=\sqrt{\frac{\gamma_{c}}{2\pi}}\sum_{n=0}^{\infty}\frac{\langle
m\vert_{b}|\tilde{n}\rangle_{b}\langle \tilde{n}|_{b}\vert
n_{0}\rangle_{b}e^{-i(m\omega_{M}+\Delta_{k})t} }
{\Delta_{k}+\delta-(n-m)\omega_{M}+i\frac{\gamma_{c}}{2}}.\label{lontimsoemiss}
\end{equation}
We point out that the long-time limit here actually refers to the
time scale of $1/\gamma_c\ll t\ll 1/\gamma_m$. During this time
duration, the single photon leaks completely out of the cavity and
the damping of the mechanical motion is negligible.

To understand Eq.~(\ref{lontimsoemiss}), we note that the mirror's
initial state $|n_{0}\rangle_{b}$ is a superposition of displaced
number states $|n_{0}\rangle_{b}=\sum_{n=0}^{\infty}(\langle
\tilde{n}|_{b}|n_{0}\rangle_{b})|\tilde{n}\rangle_{b}$. The
transition $|\tilde{n}\rangle_{b}\rightarrow |m\rangle_{b}$ can
occur in the process of cavity photon leakage
$|1\rangle_{a}\rightarrow|0\rangle_{a}$. For the process
$|1\rangle_{a}|\tilde{n}\rangle_{b}\rightarrow
|0\rangle_{a}|m\rangle_{b}$, the frequency of the emitted photon is
governed by the resonance condition [see Fig.~\ref{setup}(b)]
\begin{equation}
\Delta_{k}=(n-m)\omega_{M}-\delta,\label{resoncondi}
\end{equation}
which is consistent with the real part of the pole in
Eq.~(\ref{lontimsoemiss}), [i.e.,
$\Delta_{k}+\delta-(n-m)\omega_{M}=0$]. The amplitude for the
process is proportional to the overlap $\langle
m|_{b}|\tilde{n}\rangle_{b}\langle \tilde{n}|_{b}\vert
n_{0}\rangle_{b}$, which can be calculated with the
relation~\cite{Oliveira1990}
\begin{eqnarray}
\langle m\vert_{b}\vert\tilde{n}\rangle_{b}=\left\{
\begin{array}{c}
\sqrt{\frac{m!}{n!}}e^{-\frac{\beta_{0}^{2}}{2}}
(-\beta_{0})^{n-m}L_{m}^{n-m}(\beta_{0}^{2}),\hspace{0.1 cm}n\geq m,\\
\sqrt{\frac{n!}{m!}}e^{-\frac{\beta_{0}^{2}}{2}}
\beta_{0}^{m-n}L_{n}^{m-n}(\beta_{0}^{2}),\hspace{0.1 cm}m>n,
\end{array}
\right.
\end{eqnarray}
where $L_{r}^{s}(x)$ is the associated Laguerre polynomial.

A useful quantity in this system is the final reservoir occupation
spectrum $S(\Delta_{k})$~\cite{Garraway2008}, which is the
probability density for finding the single photon in the $k$th mode
of these outside fields. When the mirror is initially in a pure
state $\sum_{n_{0}=0}^{\infty}C_{n_{0}}|n_{0}\rangle_{b}$ or a mixed
state $\sum_{n_{0}=0}^{\infty}P_{n_{0}}|n_{0}\rangle_{b}\langle
n_{0}|_{b}$, the single-photon emission spectra are, respectively,
given by
\begin{subequations}
\label{av-nk}
\begin{align}
S(\Delta_{k})&=\sum_{m=0}^{\infty} \left|\sum_{n_{0}=0}^{\infty}
C_{n_{0}}B_{n_{0},m,k}(\infty)\right|^{2},\label{av-nk1}\\
S(\Delta_{k})&=\sum_{m=0}^{\infty}\sum_{n_{0}=0}^{\infty}P_{n_{0}}
\left|B_{n_{0},m,k}(\infty)\right|^{2}.\label{av-nk2}
\end{align}
\end{subequations}
For a thermal equilibrium state $\rho^{th}_{b}$ with thermal phonon
number $\bar{n}_{b}$, we have
$P_{n_{0}}=\bar{n}_{b}^{n_{0}}/(\bar{n}_{b}+1)^{n_{0}+1}$.
\begin{figure}[tbp]
\center
\includegraphics[bb=38 2 523 410, width=3.3 in]{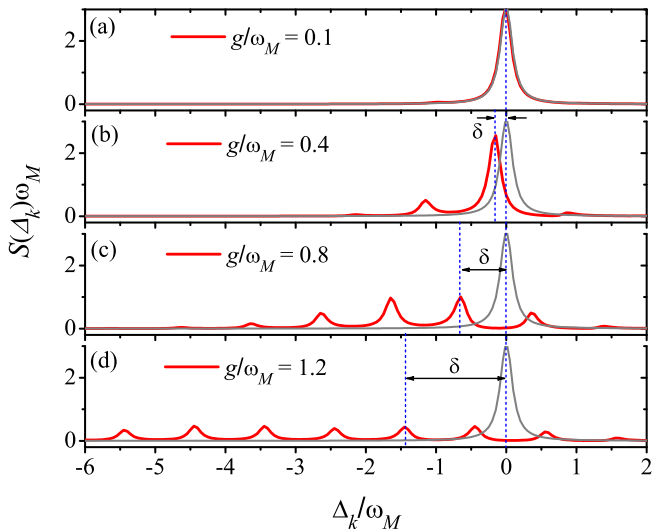}
\caption{(Color online) Spectrum $S(\Delta_{k})$ (red curves) of
single-photon emission versus $\Delta_{k}$ for various $g$. The
mirror's initial state is $|0\rangle_{b}$ and
$\gamma_{c}/\omega_{M}=0.2$. The dark gray curve is a Lorentzian
spectrum for comparison.} \label{emiorivariousg}
\end{figure}

To observe the spectral signatures of the coupling strength $g$, in
Fig.~\ref{emiorivariousg} we plot the spectrum $S(\Delta_{k})$ of
single-photon emission as a function of the photon frequency
$\Delta_{k}$ (in rotating picture) for various values of $g$ when
the mirror is initially prepared in ground state $|0\rangle_{b}$.
Figure~\ref{emiorivariousg}(a) is plotted for $g<\gamma_{c}$, there
are no phonon sidebands. However, spectral peaks of phonon sidebands
become visible when $g>\gamma_c$ [Figs.~\ref{emiorivariousg}(b) to
\ref{emiorivariousg}(d)]. Such a condition can be understood by
examining the overlap of Lorentizians in Eq.~(\ref{lontimsoemiss}).
To resolve a peak in the spectrum, the peak height should be higher
than the tail of its neighboring Lorentizian. This requires
$g>\gamma_c$ in the resolved sideband regime~\cite{remark}.
\begin{figure}[tbp]
\center
\includegraphics[bb=38 2 523 410, width=3.3 in]{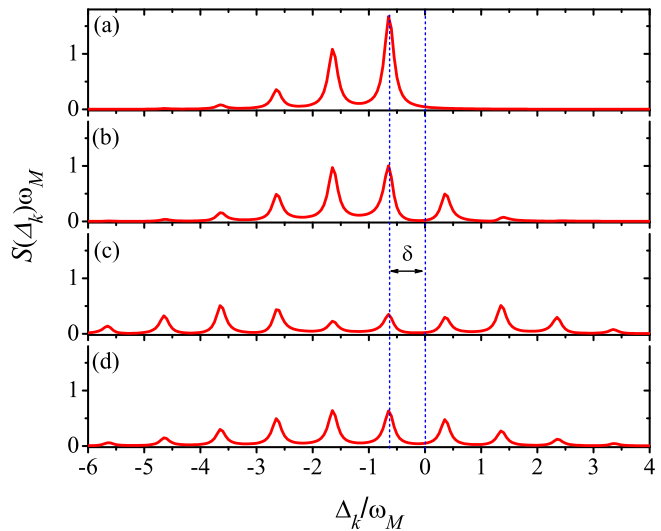}
\caption{(Color online) Spectrum $S(\Delta_{k})$ of single-photon
emission versus $\Delta_{k}$ when the mirror is initially in (a)
displaced ground state $|\tilde{0}\rangle_{b}$, (b) ground state
$|0\rangle_{b}$, (c) coherent state $|\beta\rangle_{b}$ with
$\beta=3$, and (d) thermal state $\rho^{th}_{b}$ with
$\bar{n}_{b}=2$. Here $g/\omega_{M}=0.8$ and
$\gamma_c/\omega_m=0.2$.} \label{emioriinistate}
\end{figure}

From Eq.~(\ref{resoncondi}) the peaks of these sidebands are located
at $\Delta_{k}=(n-m)\omega_{M}-\delta$. Therefore there is always a
peak at $\Delta_k=-\delta$ (for $n=m$), which is shown in
Figs.~\ref{emiorivariousg}(b) to \ref{emiorivariousg}(d). In
particular for $g<\omega_M$, the $\Delta_k=-\delta$ peak appears at
the first (labeled from right to left in the region $\Delta_{k}<0$)
peak in the red sideband. We remark that the photon-state frequency
shift $\delta=g^{2}/\omega_{M}$ can be resolved from the Lorentzian
spectrum when $\delta>\gamma_{c}$, which requires
$g>\sqrt{\omega_{M}\gamma_{c}}$.

To illustrate how the spectrum $S(\Delta_{k})$ depends on the
mirror's initial state, we plot in Fig.~\ref{emioriinistate} the
emission spectrum $S(\Delta_{k})$ as a function of the photon
frequency $\Delta_{k}$. Four different initial states are considered
here: displaced ground state $|\tilde{0}\rangle_{b}$, ground state
$|0\rangle_{b}$, coherent state $|\beta\rangle_{b}$, and thermal
state $\rho^{th}_{b}$. When the mirror is initially in
$|\tilde{0}\rangle_{b}$, the photon leaves the cavity with a
frequency smaller than $\omega_{c}$, and so there are only peaks at
the red sideband. In contrast, for the cases of $|0\rangle_{b}$,
$|\beta\rangle_{b}$, and $\rho^{th}_{b}$, the mirror in the
displacement representation have some excited-state populations
because of $|n_{0}\rangle_{b}=\sum_{n=0}^{\infty}(\langle
\tilde{n}|_{b}|n_{0}\rangle_{b})|\tilde{n}\rangle_{b}$. Consequently
there will be some probabilities for the single photon absorbing the
phonon's energy and leaving the cavity with a frequency larger than
$\omega_{c}$, (i.e., there are some peaks at the blue sideband).
Figure~\ref{emioriinistate} also indicates that the number of these
peaks depends on the initially contributing phonon distribution in
the mirror. The wider the contributing phonon distribution is, the
more the peak's number becomes.

\emph{Single-photon scattering.}---Now we turn to the photon
scattering problem in which a single photon is injected into the
cavity. We assume that initially the cavity is in a vacuum, the
single photon in a Lorentzian wave packet~\cite{Liao2010}, and the
mirror in number state $|n_{0}\rangle_{b}$, that is,
\begin{eqnarray}
A_{m}(0)=0,\hspace{0.5
cm}B_{m,k}(0)=\sqrt{\frac{\epsilon}{\pi}}\frac{\delta_{m,n_{0}}}
{\Delta_{k}-\Delta_{0}+i\epsilon},
\end{eqnarray}
where $\Delta_{0}$
and $\epsilon$ are the detuning and spectral width of the photon,
respectively. The scattering solution of the system is obtained as
\begin{figure}[tbp]
\center
\includegraphics[bb=38 2 523 410, width=3.3 in]{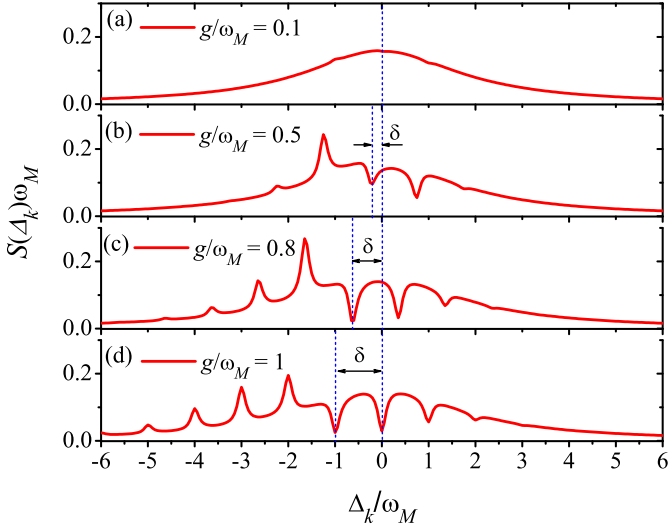}
\caption{(Color online) Spectrum $S(\Delta_{k})$ of single-photon
scattering versus $\Delta_{k}$ for various $g$. The initial state of
the mirror is $|0\rangle_{b}$. Other parameters are set as
$\gamma_{c}/\omega_{M}=0.2$, $\epsilon/\omega_{M}=2$, and
$\Delta_{0}/\omega_{M}=0$.} \label{widebandscat}
\end{figure}
\begin{subequations}
\label{bmkt}
\begin{align}
A_{n_{0},m}(t)=&\sqrt{\frac{\epsilon}{\pi}}\frac{-2\pi\xi\langle
\tilde{m}\vert_{b}\vert
n_{0}\rangle_{b}}{\epsilon-\frac{\gamma_{c}}{2}+i[(n_{0}-m)\omega_{M}+\Delta
_{0}+\delta]}\nonumber\\
&\times\left(e^{-[\frac{\gamma_{c}}{2}+i(m\omega_{M}
-\delta)]t}-e^{-[\epsilon+i(n_{0}\omega_{M}+\Delta_{0})]t}\right),\label{bmkt1}\\
B_{n_{0},m,k}(t)=&\sqrt{\frac{\epsilon}{\pi}}\frac{1}
{\Delta_{k}-\Delta_{0}+i\epsilon
}\delta_{m,n_{0}}e^{-i(m\omega_{M}+\Delta_{k})t}\notag \\
&+\sum_{n=0}^{\infty}\frac{i\gamma_{c}\sqrt{\epsilon/\pi}\langle
m\vert_{b}\vert\tilde{n}\rangle_{b}\langle\tilde{n}\vert_{b}\vert
n_{0}\rangle_{b}}
{\epsilon-\frac{\gamma_{c}}{2}+i[(n_{0}-n)\omega_{M}+\Delta_{0}+\delta]}\nonumber\\
&\times\frac{\left(e^{-i(m\omega_{M}
+\Delta_{k})t}-e^{-\left[\frac{\gamma_{c}}{2}+i(n\omega_{M}
-\delta)\right]t}\right)}{\frac{\gamma_{c}}{2}+i[(n-m)\omega_{M}-\delta-\Delta_{k}]}\notag\\
&-\sum_{n=0}^{\infty}\frac{i\gamma_{c}\sqrt{\epsilon/\pi}\langle
m\vert_{b}\vert\tilde{n}\rangle_{b}\langle \tilde{n}\vert_{b}\vert
n_{0}\rangle_{b}}{\epsilon-\frac{\gamma_{c}}{2}+i[
(n_{0}-n)\omega_{M}+\Delta_{0}+\delta]}\nonumber\\
&\times\frac{\left(e^{-i(m\omega_{M}
+\Delta_{k})t}-e^{-[\epsilon+i(n_{0}\omega_{M}
+\Delta_{0})]t}\right)}{\epsilon
+i[(n_{0}-m)\omega_{M}+\Delta_{0}-\Delta_{k}]}.\label{bmkt2}
\end{align}
\end{subequations}
In the long-time limit, $A_{n_{0},m}(\infty)=0$ and
\begin{eqnarray}
B_{n_{0},m,k}(\infty)&=&\sqrt{\frac{\epsilon}{\pi
}}\frac{e^{-i(m\omega_{M}+\Delta_{k})t}}{\Delta_{k}-\Delta_{0}+i\epsilon
}\delta_{m,n_{0}}\nonumber\\
&&-\sqrt{\frac{\epsilon}{\pi
}}\frac{i\gamma_{c}}{\Delta_{k}-[\Delta_{0}+(n_{0}-m)\omega_{M}]+i\epsilon}\nonumber\\
&&\times\sum_{n=0}^{\infty}\frac{\langle m\vert_{b}\vert
\tilde{n}\rangle_{b}\langle\tilde{n}\vert_{b}\vert
n_{0}\rangle_{b}e^{-i(m\omega_{M}+\Delta_{k})t}}{\Delta_{k}+\delta-(n-m)\omega_{M}+i\frac{\gamma_{c}}{2}}.\label{bmkscattering}
\end{eqnarray}
The first term in Eq.~(\ref{bmkscattering}) corresponds to the
process in which the photon is reflected by the fixed cavity mirror
without entering the cavity. The second term in
Eq.~(\ref{bmkscattering}) comes from the interaction process after
the single photon entering the cavity. Note that the summation part
in the second term of Eq.~(\ref{bmkscattering}) has the same form as
that appearing in single-photon emission process discussed above.

Let us first consider the case of $\epsilon/\omega_{M}>1$. We plot
in Fig.~\ref{widebandscat} the spectrum $S(\Delta_{k})$ of
single-photon scattering versus the photon detuning $\Delta_{k}$ for
various $g$ when the mirror is initially prepared in ground state
$|0\rangle_{b}$. It can be seen from Fig.~\ref{widebandscat} that
phonon sidebands appear in the spectrum when $g>\gamma_{c}$. In
addition, the sideband peaks in the spectrum can also be used to
characterize the coupling strength $g$ through the frequency shift
$\delta$, similar to Fig.~\ref{emiorivariousg}. It is interesting
that these sidebands in the spectrum show both peaks and dips. The
reason for these tips is quantum interference between the direct
photon reflection channel and the scattering channel relating with
the mirror's final state $|0\rangle_{b}$, which are, respectively,
the first and second terms in $B_{0,0,k}(\infty)$ of
Eq.~(\ref{bmkscattering}).
\begin{figure}[tbp]
\center
\includegraphics[bb=38 2 523 410, width=3.3 in]{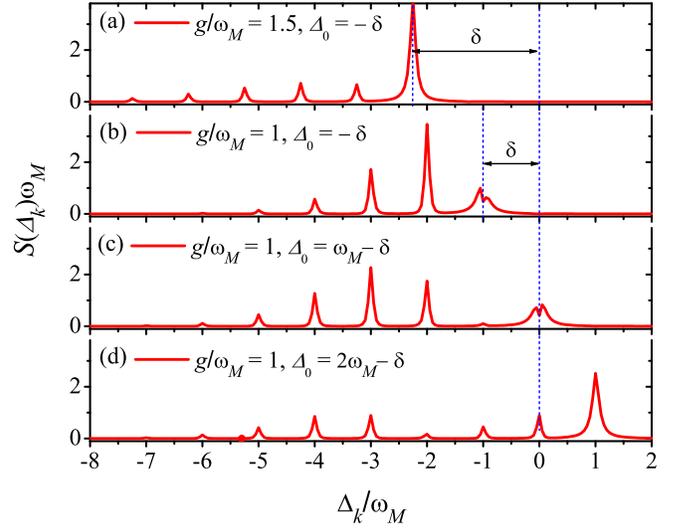}
\caption{(Color online) Spectrum $S(\Delta_{k})$ of single-photon
scattering versus $\Delta_{k}$ for various $g$ and $\Delta_{0}$. The
initial state of the mirror is $|0\rangle_{b}$. Here
$\gamma_c/\omega_m=0.2$ and $\epsilon/\omega_{M}=0.05$.}
\label{exactdriscatt}
\end{figure}

We next consider the case in which the frequency of the incident
photon is nearly monochromatic, (i.e., $\epsilon\ll\gamma_{c}$). The
motivation for studying the case is to see the resonant transition
processes by tuning the incident photon frequency $\Delta_{0}$. In
Figs.~\ref{exactdriscatt}(a) and~\ref{exactdriscatt}(b), we choose
$\Delta_{0}=-\delta$ such that the incident photon will excite
resonantly the transition
$|0\rangle_{a}|0\rangle_{b}\rightarrow|1\rangle_{a}|\tilde{0}\rangle_{b}$.
A subsequent cavity photon decay will induce transitions
$|1\rangle_{a}|\tilde{0}\rangle_{b}\rightarrow|0\rangle_{a}|n\rangle_{b}$
$(n=0,1,2,\cdots)$. As a result, there are only some peaks at the
red sideband, and the first sideband peak measured from
$\Delta_{k}=0$ is located at $\Delta_{k}=-\delta$. On the other
hand, Figs.~\ref{exactdriscatt}(b) to \ref{exactdriscatt}(d) provide
a comparison of the scattering spectrum when the incident photon
frequencies are: $\Delta_{0}=-\delta$, $\omega_{M}-\delta$, and
$2\omega_{M}-\delta$ such that the photon resonantly excites the
system from states $|0\rangle_{a}|0\rangle_{b}$ to
$|1\rangle_{a}|\tilde{0}\rangle_{b}$,
$|1\rangle_{a}|\tilde{1}\rangle_{b}$, and
$|1\rangle_{a}|\tilde{2}\rangle_{b}$, respectively. When the photon
leaks out of the cavity, the mirror will transition from
$|\tilde{0}\rangle_{b}$, $|\tilde{1}\rangle_{b}$, and
$|\tilde{2}\rangle_{b}$ to $|n\rangle_{b}$ $(n=0,1,2,\cdots)$,
respectively. Then according to Fig.~\ref{setup}(b) the
maximal-frequency sideband peak should be located at
$\Delta_{k}=-\delta$, $\omega_{M}-\delta$, and $2\omega_{M}-\delta$,
as illustrated in Fig.~\ref{exactdriscatt}.

In conclusion, we have calculated analytically the spectrum of
single-photon emission and scattering in a cavity optomechanical
system. We have also indicated the connection between the spectral
features and the interaction strength $g$ of radiation pressure per
photon. In the resolved sideband regime $\omega_{M}\gg\gamma_{c}$,
the phonon sidebands are visible when $g>\gamma_{c}$, while the
condition for resolving the photon-state frequency shift $\delta$ is
$g>\sqrt{\omega_{M}\gamma_{c}}$.

This work is partially supported by a grant from the Research Grants
Council of Hong Kong, Special Administrative Region of China
(Project No.~CUHK401810).


\begin{thebibliography}{99}
\bibitem{Kippenberg2008}     T. J. Kippenberg and K. J. Vahala, Science \textbf{321}, 1172 (2008).
\bibitem{Marquardt2009}      F. Marquardt and S. M. Girvin, Physics \textbf{2}, 40 (2009).

\bibitem{Gupta2007}          S. Gupta, K. L. Moore, K. W. Murch, and D. M. Stamper-Kurn, Phys. Rev. Lett. \textbf{99}, 213601 (2007).
\bibitem{Brennecke2008}      F. Brennecke, S. Ritter, T. Donner, and T. Esslinger, Science \textbf{322}, 235 (2008).
\bibitem{Eichenfield2009}    M. Eichenfield, J. Chan, R. M. Camacho, K. J. Vahala, and O. Painter, Nature (London) \textbf{462}, 78 (2009).

\bibitem{Mancini1997}        S. Mancini, V. I. Manko, and P. Tombesi, Phys. Rev. A \textbf{55}, 3042 (1997).
\bibitem{Bose1997}           S. Bose, K. Jacobs, and P. L. Knight, Phys. Rev. A \textbf{56}, 4175 (1997).
\bibitem{Bouwmeester2003}    W. Marshall, C. Simon, R. Penrose, and D. Bouwmeester, Phys. Rev. Lett. \textbf{91}, 130401 (2003).

\bibitem{Rabl2011}           P. Rabl, Phys. Rev. Lett. \textbf{107}, 063601 (2011).
\bibitem{Nunnenkamp2011}     A. Nunnenkamp, K. B{\o}rkje, and S. M. Girvin, Phys. Rev. Lett. \textbf{107}, 063602 (2011).
\bibitem{Chenp2011}          T. Hong, H. Yang, H. Miao, and Y. Chen, e-print arXiv:1110.3348.

\bibitem{Oliveira1990}       F. A. M. de Oliveira, M. S. Kim, P. L. Knight, and V. Buzek, Phys. Rev. A \textbf{41}, 2645 (1990).
\bibitem{Garraway2008}       I. E. Linington and B. M. Garraway, Phys. Rev. A \textbf{77}, 033831 (2008).

\bibitem{remark}             For example, we consider the case of $g/\omega_{M}\ll1$. In the resolved sideband regime
                             $\omega_{M}\gg \gamma_{c}$ and under the initial state $|0\rangle_{b}$, the main peak of
                             the spectrum is approximately a Lorentizian function $S_{L}(\Delta_{k})\approx\frac{\gamma_{c}}{2\pi}\frac{1}{(\Delta
                             _{k}+\delta)^{2}+(\gamma_{c}/2)^{2}}$. From Eqs.~(\ref{lontimsoemiss}) and (\ref{av-nk1}) the height of the second red sideband
                             can be approximated as
                             $S(-\omega_{M}-\delta)\approx\frac{\gamma_{c}}{2\pi\omega^{2}_{M}}\left(1+\frac{4g^{2}}{\gamma^{2}_{c}}\right)$
                             up to second order of $g/\omega_{M}$. Then the requirement
                             $\frac{S(-\omega_{M}-\delta)}{S_{L}(-\omega_{M}-\delta)}\approx1+4g^{2}/\gamma^{2}_{c}\gg1$
                             leads to the condition $g\gg\gamma_{c}$.
\bibitem{Liao2010}           J. Q. Liao and C. K. Law, Phys. Rev. A \textbf{82}, 053836 (2010).
\end{thebibliography}
\end{document}